\documentclass[12pt]{article}
\usepackage{epsfig}
\begin{document}
\title{A renormalization group study of the three-color Ashkin-Teller model on
          a Wheatstone hierarchical lattice}
\author{R. Teodoro, C.G. Bezerra, A.M. Mariz and F.A. da Costa \\ 
Departamento de F\'\i sica Te\'orica e Experimental,\\ Universidade Federal do Rio  Grande do Norte, \\
Caixa Postal 1641, \\
Natal-RN 59072-970, Brazil}

\maketitle
\begin{abstract}
We have investigated the three-color Ashkin-Teller model (3AT), on the Wheatstone bridge hierarchical lattice, by means of
a Migdal-Kadanoff renormalization group approach. We have obtained the exact recursion relations for the
renormalized couplings, which have been used to investigate the phase diagram
and to study the corresponding critical points. The phase diagram, represented in terms of the dual transmissivity vector, presents four magnetic phases and nine critical points. We
have also numerically calculated the correlation length ($\nu_T$) and crossover ($\phi$)
critical exponents, which show that seven of the critical points are in the Potts model universality class ($q=2$, $4$ e $8$). The remaining critical points are
in a universality class which may belong to the Baxter's line. Our results are exact on the hierarchical lattice used in the present work and the phase diagram can be considered as an approximation
to more realistic Bravais lattices.

\end{abstract}


\section{Introduction}
The N-color Ashkin-Teller (AT) model was introduced in order to investigate some aspects related to the non-universal critical behavior presented by the usual two-color AT model (2AT) \cite{gw1981}. Such non-universal behavior manifests itself in the two-dimensional 2AT model which exhibits a line of varying critical exponents \cite{baxter}.
However, it was found that the non-universal behavior is a peculiar feature of the 2AT model. In Ref.\ \cite{gw1981} a series of techniques, such as first-order perturbation expansion, mean-field theory and Monte Carlo simulations, were used to investigate the three-color AT (3AT) phase diagram in two and three dimensions. Since then, the 3AT model has attracted some attention, mainly due to the richness of its phase diagram \cite{gs1986,mf1988,rf1996,fcc2003}. In a recent study, the phase diagram of the 3AT model was investigated by means of the exact Migdal-Kadanoff real space renormalization group treatment on a diamond hierarchical lattice \cite{pi2008}.  In this work we perform a real space renormalization group analysis of the 3AT model on the Wheatstone hierarchical lattice \cite{tm1996}. Hierarchical lattices are constructed recursively in such a way that a large class of models can be solved exactly through a decimation procedure \emph{\`a la} Migdal-Kadanoff real space renormalization group scheme \cite{tm1996,kg1981}.

In the present work we use a hierarchical lattice recursively constructed from an initial unit - the Wheatstone bridge - and a geometrical substitution rule as depicted in Fig.\ 1. Due to the lattice structure, we are restricted to consider only the ferromagnetic version of the 3AT model under investigation. Our results are exact in such a lattice and the phase diagrams may be considered as approximations to more realistic Bravais lattices.
The plan of this article is as follows. In Section 2, we introduce the model and obtain the exact recursion relations for the renormalized couplings.  In Section 3 we analyse the structure of the fixed points and determine the critical behavior of the model and the corresponding phase diagram. Finally, in Section 4 we summarize our main findings.

\section{The model and the renormalization group relations}

Let us consider the 3AT model consisting of three Ising systems coupled by both four- and six-spin interaction terms. The Hamiltonian is given by

\begin{eqnarray}
\beta {\mathcal{H}}& = & -\sum_{\langle i,j \rangle}[K_2(\sigma_i\sigma_j+\theta_i\theta_j+\tau_i\tau_j)+K_4(\sigma_i\sigma_j\theta_i\theta_j+\sigma_i\sigma_j\tau_i\tau_j+\theta_i\theta_j\tau_i\tau_j) \nonumber \\  & & + K_6(\sigma_i\sigma_j\theta_i\theta_j\tau_i\tau_j)],\label{3colors}
\end{eqnarray}

\noindent where $\beta = 1/k_{B}T$, the sum is over the nearest-neighbor pairs of sites $i$ and $j$ and each spin variable ($\sigma$, $\theta$ and $\tau$) assumes one of two possible values independently: $\sigma=\pm1$, $\theta=\pm1$ and $\tau=\pm1$. Also, $K_2=J_2/k_BT$, $K_4=J_4/k_BT$ and $K_6=J_6/k_BT$ represent dimensionless coupling constants.

The 3AT model described above will be studied on the hierarchical lattice generated as shown in Fig.\ 1. For this purpose we will employ a real space renormalization group approach which is known to be exact for hierarchical lattices and to give a good qualitative approximation for phase diagrams in real Bravais lattices \cite{tm1996,kg1981}. To treat the Hamiltonian of Eq. \ref{3colors} we apply the renormalization group transformation indicated in Fig.\ 2. The renormalization group recursive relations are obtained in two basic steps: (i) by the decimation of the spin variables associated with the intermediate sites 3 and 4; (ii) followed by the requirement that the partition function should be preserved.

As we have already mentioned in the Introduction, this kind of lattice becomes very suitable to investigate ferromagnetic order. For instance, in the case of the Potts model, it yields
the exact transition temperature for the square lattice and good approximations to the critical exponents wherever it applies. Therefore, we will consider only stable ferromagnetic phases. This restriction impose some constraints that must be obeyed by the coupling constants as inequalities:
\begin{equation}
 K_2 + K_4 \geq 0,\qquad 3K_2 + K_6 \geq 0\quad\mbox{and}\quad K_2 + 2K_4 + K_6 \geq 0.
\end{equation}

From now on we will make use of the transmissivity vector and its dual \cite{at1982}, as it was done in a recent work \cite{pi2008}. These vectors are very useful since they simplify some intermediate calculations and become very convenient to represent the phase diagram. The transmissivity vector is defined by its components \cite{at1982}
\begin{equation}
t_2=\frac{1+e^{(-2K_2-4K_4-2K_6)}-e^{(-4K_2-4K_4)}-e^{(-6K_2-2K_6)}}{1+3e^{(-2K_2-4K_4-2K_6)}+3e^{(-4K_2-4K_4)}+e^{(-6K_2-2K_6)}},
\end{equation}
\begin{equation}
 t_4=\frac{1-e^{(-2K_2-4K_4-2K_6)}-e^{(-4K_2-4K_4)}+e^{(-6K_2-2K_6)}}{1+3e^{(-2K_2-4K_4-2K_6)}+3e^{(-4K_2-4K_4)}+e^{(-6K_2-2K_6)}} ,
\end{equation}

\noindent and
\begin{equation}
 t_6=\frac{1-3e^{(-2K_2-4K_4-2K_6)}+3e^{(-4K_2-4K_4)}-e^{(-6K_2-2K_6)}}{1+3e^{(-2K_2-4K_4-2K_6)}+3e^{(-4K_2-4K_4)}+e^{(-6K_2-2K_6)}} ,
\end{equation}

\noindent
whereas the dual transmissivity vector has components given by
\begin{equation}
 \bar{t}_2 = \frac{1+t_2-t_4-t_6}{1+3t_2+3t_4+t_6}=e^{(-2K_2-4K_4-2K_6)} ,
\end{equation}
\begin{equation}
 \bar{t}_4 = \frac{1-t_2-t_4+t_6}{1+3t_2+3t_4+t_6}=e^{(-4K_2-4K_4)} ,
\end{equation}

\noindent and
\begin{equation}
 \bar{t}_6 = \frac{1-3t_2+3t_4-t_6}{1+3t_2+3t_4+t_6}=e^{(-6K_2-2K_6)}.
\end{equation}

After some calculations along the lines described in Refs.\ \cite{pi2008} and \cite{tm1996}, we obtain the following recursion relations

\begin{equation}
 \bar{t}^{~'}_2 = \frac{N_1}{D},\label{t2}
\end{equation}
\begin{equation}
 \bar{t}^{~'}_4 = \frac{N_2}{D}\label{t4} ,
\end{equation}
\noindent and
\begin{equation}
  \bar{t}^{~'}_6 = \frac{N_3}{D},\label{t6}
\end{equation}

\noindent where  
 \newline
\begin{eqnarray} \label{eqdn}
D&=& \bar{t}_6^{~4} + 2  \bar{t}_6^{~3} + 6 \bar{t}_2  \bar{t}_4^{~2}  \bar{t}_6^{~2} + 6  \bar{t}_2^{~2} \bar{t}_4 \bar{t}_6^{~2} + 6 \bar{t}_2^{~2} \bar{t}_4^{~2} \bar{t}_6 \nonumber \\
& & + ~6 \bar{t}_4^{~5} + 3 \bar{t}_4^{~4} + 6 \bar{t}_4^{~3} + 12 \bar{t}_2^{~3} \bar{t}_4^{~2} + 6 \bar{t}_2^{~4} \bar{t}_4 \nonumber \\
& & + ~3 \bar{t}_2^{~4} + 6 \bar{t}_2^{~3} + 1,
\end{eqnarray}
\begin{eqnarray}
N_1 &=& 2( \bar{t}_2 \bar{t}_4^{~2} \bar{t}_6^{~2} + \bar{t}_4^{~2} \bar{t}_6^{~2} + 2 \bar{t}_2 \bar{t}_4 \bar{t}_6^{~2} + 4 \bar{t}_2 \bar{t}_4^{~3} \bar{t}_6 \nonumber \\
& & + ~6 \bar{t}_2^{~2} \bar{t}_4^{~2} \bar{t}_6 + 2 \bar{t}_2 \bar{t}_4^{~2} \bar{t}_6 + 2 \bar{t}_2^{~2} \bar{t}_4^{~3} + 2 \bar{t}_2^{~3} \bar{t}_4^{~2} \nonumber \\
& & + ~6 \bar{t}_2^{~2} \bar{t}_4^{~2} + 4 \bar{t}_2^{~3} \bar{t}_4 + \bar{t}_2^{~3} + \bar{t}_2^{~2}),
\end{eqnarray}
\begin{eqnarray}
N_2 &=& 2( \bar{t}_2^{~2} \bar{t}_4 \bar{t}_6^{~2} + 2 \bar{t}_2 \bar{t}_4 \bar{t}_6^{~2} + \bar{t}_2^{~2} \bar{t}_6^{~2} + 6 \bar{t}_2^{~2} \bar{t}_4^{~2} \bar{t}_6 \nonumber \\
& & + ~4 \bar{t}_2^{~3} \bar{t}_4 \bar{t}_6 + 2 \bar{t}_2^{~2} \bar{t}_4 \bar{t}_6 + \bar{t}_4^{~5} + 5 \bar{t}_4^{~4} + \bar{t}_4^{~3}  \nonumber \\
& & + ~2 \bar{t}_2^{~3} \bar{t}_4^{~2} + \bar{t}_4^{~2} + \bar{t}_2^{~4} \bar{t}_4 + 4 \bar{t}_2^{~3} \bar{t}_4 + \bar{t}_2^{~4}),
\end{eqnarray}
\noindent
and
\begin{eqnarray} \label{eqn3}
N_3 &=& 2( \bar{t}_6^{~3} + \bar{t}_6^{~2} + 3 \bar{t}_2^{~2} \bar{t}_4^{~2} \bar{t}_6 + 6 \bar{t}_2 \bar{t}_4^{~2} \bar{t}_6 \nonumber \\
& & + ~6 \bar{t}_2^{~2} \bar{t}_4 \bar{t}_6 + 6 \bar{t}_2^{~2} \bar{t}_4^{~3} + 6 \bar{t}_2^{~3} \bar{t}_4^{~2}  + 3 \bar{t}_2^{~2} \bar{t}_4^{~2} ).
\end{eqnarray}

The recursion relations given by Eqs. \ref{t2}-\ref{t6} completely determine the critical exponents as well as the phase diagram for the 3AT model. For instance, their fixed points and the renormalization flows allow us to obtain the phase diagram. On the other hand, the critical exponents are obtained from the linearization around the fixed points. This will be discussed in the next section.

\section{Results and the phase diagram}

Using Eqs. \ref{t2}-\ref{t6} we were able to numerically determine the existence of four distinct phases separated by two-dimensional critical surfaces. The phases are characterized as follows (see also Fig.\ 3)

\begin{itemize}
    \item   \textbf{paramagnetic} (\textbf{P}) $\rightarrow$ corresponding to the trivial fixed point $\bar{t}_2 = \bar{t}_4 = \bar{t}_6 = 1$ and in which $\langle\sigma\rangle=\langle\theta\rangle=\langle\tau\rangle=0$, $\langle\sigma\theta\rangle=\langle\sigma\tau\rangle=\langle\theta\tau\rangle=0$ and $\langle\sigma\theta\tau\rangle=0$; \end{itemize}
\begin{itemize}
	 \item \textbf{ferromagnetic} (\textbf{F})  $\rightarrow$ corresponding to the trivial fixed point $\bar{t}_2 = \bar{t}_4 = \bar{t}_6 = 0$ and in which $\langle\sigma\rangle\neq0$, $\langle\theta\rangle\neq0$, $\langle\tau\rangle\neq0$, $\langle\sigma\theta\rangle\neq0$, $\langle\sigma\tau\rangle\neq0$, $\langle\theta\tau\rangle\neq0$ and $\langle\sigma\theta\tau\rangle\neq0$;
\end{itemize}
\begin{itemize}
	 \item \textbf{intermediate-one} ($\mathbf{F}_1$)  $\rightarrow$ corresponding to the trivial fixed point $\bar{t}_4 = 1$, $\bar{t}_2 = \bar{t}_6 = 0$ and in which $\langle\sigma\rangle=\langle\theta\rangle=\langle\tau\rangle=0$, $\langle\sigma\theta\rangle=\langle\sigma\tau\rangle=\langle\theta\tau\rangle=0$ and $\langle\sigma\theta\tau\rangle\neq0$;
\end{itemize}
\begin{itemize}
	\item \textbf{intermediate-two }($\mathbf{F}_2$) $\rightarrow$ corresponding to the trivial fixed point $\bar{t}_6 = 1$, $\bar{t}_2 = \bar{t}_4 = 0$ and  in which $\langle\sigma\rangle=\langle\theta\rangle=\langle\tau\rangle=0$, $\langle\sigma\theta\rangle\neq0$, $\langle\sigma\tau\rangle\neq0$, $\langle\theta\tau\rangle\neq0$ and $\langle\sigma\theta\tau\rangle=0$.
\end{itemize}

\vspace{0.5cm}

\begin{table}[htb]\caption{Location of the trivial fixed points and their respective magnetic phases.}\label{fixedpoints}
\begin{center}
   \begin{tabular*}{\textwidth}{@{\extracolsep{\fill}}lll}
      \hline
      Trivial fixed point & ($\bar{t}_2$, $\bar{t}_4$, $\bar{t}_6$) & Magnetic phase \\
      \hline \hline
      $\mathbf{F}$ & $(0, 0, 0)$ & Ferromagnetic \\
      $\mathbf{F}_1$ & $(0, 1, 0)$ & Intermediate-one \\
      $\mathbf{F}_2$ & $(0, 0, 1)$ & Intermediate-two \\
      $\mathbf{P}$ & $(1, 1, 1)$ & Paramagnetic \\
      \hline
   \end{tabular*}
   \end{center}
  \end{table}

\vspace{0.5cm}

Each phase is characterized by its corresponding stable fixed point which are summarized in Table 1. We have also found nine additional non-trivial fixed points, identified as the critical ones. The linearization around a non-trivial fixed point allows us to determine the associated critical exponents as

\begin{equation}
	\nu_T={\ln b}/{\ln\lambda_1}
\end{equation}
and
\begin{equation}
\phi={\ln\lambda_1}/{\ln\lambda_2}.
\end{equation}

\noindent
Here $\lambda_1 > \lambda_2$ are the two largest eigenvalues of the stability matrix and $b$ is the scaling factor of the Wheatstone bridge. The critical points are presented in Table 2 with their respective critical exponents. We also note that except for the last two points the exact location of the critical points as well as the exact values for the critical exponents are given in closed form.

The fixed points $I_1$, $I_2$ and $I_3$ have exactly one relevant eigenvalue and they belong to the Ising universality class (two-state Potts model). The unstable fixed point $I_1$ is located on the surface which separates the $\mathbf{F}$ and $\mathbf{F}_2$ phases. In particular, it lies on the line given by $\bar{t}_2 = \bar{t}_4 = 0$ which renormalizes onto itself -- such kind of line will be called self-dual. We note also that this self-dual line connects two attractors -- $\mathbf{F}$ and $\mathbf{F}_2$ -- which corresponds to the ferromagnetic and intermediate-two phases, respectively. Similarly, the unstable fixed point $I_2$ is located on the boundary between the $\mathbf{P}$ and $\mathbf{F}_1$ phases and lies on the self-dual line characterized by $\bar{t}_2 = \bar{t}_6$ and $\bar{t}_4 = 1$. This line connects the $\mathbf{P}$ and $\mathbf{F}_1$ attractors which represent, respectively, the paramagnetic and intermediate-one phases. Finally, $I_3$ is located on the boundary between the ferromagnetic and the paramagnetic phases, and it is an unstable fixed point on the self-dual line $\bar{t}_4 = \bar{t}_2^{ ~2}$, $ \bar{t}_6 =\bar{t}_2^{~3}$ which connects the attractors $\mathbf{F}$ and $\mathbf{P}$.

From Table 2 we also note that the three fixed points labeled $P_4^{(1)}$, $P_4^{(2)}$ and $P_4^{(3)}$ correspond to the ferromagnetic phase of the 4-state Potts model. The unstable fixed point $P_4^{(1)}$ is located on the boundary between the $\mathbf{F}$ and  $\mathbf{F}_1$ phases and lies on the self-dual line $\bar{t}_2 = \bar{t}_6 = 0$ which connects the attractors $\mathbf{F}$ and $\mathbf{F}_1$. The unstable fixed point $P_4^{(2)}$ is located on the boundary between the paramagnetic and the intermediate-two phases along the self-dual line given by $\bar{t}_2 = \bar{t}_4$ and $\bar{t}_6 = 1$ which connects the attractors $\mathbf{P}$ and $\mathbf{F}_2$. Finally, the unstable fixed point $P_4^{(3)}$ is a multicritical point where all the four phases merge together and is located on the self-dual surface given by $\bar{t}_2 = \bar{t}_4 \cdot \bar{t}_6$, rightly at the intersection of two self-dual lines. One of such lines connects the critical points $I_1$ and $I_2$, while the other line connects the critical points $P_4^{(1)}$ and $P_4^{(2)}$. Both $P_4^{(1)}$ and $P_4^{(2)}$ have one relevant eigenvalue, whereas $P_4^{(3)}$ has two relevant eigenvalues.

The fixed point $P_8$ belongs to the 8-state Potts universality class and lies on the intersection of two self-dual planes, $\bar{t}_2 = \bar{t}_4$ and $\bar{t}_2 = \bar{t}_6$, i.e., it is located on the self-dual line $\bar{t}_2 = \bar{t}_4 = \bar{t}_6$ which connects the attractors $\mathbf{F}$ and $\mathbf{P}$. This fixed point is completely unstable and has three relevant eigenvalues, once $\lambda_1 > \lambda_2 =\lambda_3$, i.e., the lowest one is doubly degenerate.

\begin{table}
\caption{Unstable fixed points and their respective eigenvalues and
critical exponents.} \label{fixedpoints2}
\begin{center}
\scalebox{0.75}{%
\begin{tabular}{ccccc}
\hline Fixed &  & &  &
 \\
point  & ($\bar{t}_2$, $\bar{t}_4$, $\bar{t}_6$) & $(\lambda_1,
\lambda_2, \lambda_3)$ & $\nu_T$ &
$\phi$ \\
\hline \hline
      $P_8$ & ($\frac{2\sqrt{2}-1}{7}$, $\frac{2\sqrt{2}-1}{7}$, $\frac{2\sqrt{2}-1}{7}$) & ($\frac{32\sqrt{2}-5}{17}$, $\frac{40-\sqrt{2}}{34}$, $\frac{40-\sqrt{2}}{34}$) & $\frac{\ln(2)}{\ln(\frac{32\sqrt{2}-5}{17})}$ & $\frac{\ln(\frac{32\sqrt{2}-5}{17})}{\ln(\frac{40-\sqrt{2}}{34})}$ \\
      $P_4^{(1)}$ & (0, $\frac{1}{3}$, 0) & ($\frac{27}{13}$, 0, 0) & $\frac{\ln(2)}{\ln(\frac{27}{13})}$ &  \\
      $P_4^{(2)}$ & ($\frac{1}{3}$, $\frac{1}{3}$, 1) & ($\frac{27}{13}$, 0, 0) & $\frac{\ln(2)}{\ln(\frac{27}{13})}$ &  \\
      $P_4^{(3)}$ & ($\frac{\sqrt{2}-1}{3}$, $\frac{1}{3}$, $\sqrt{2}-1$) & ($\frac{27}{13}$, $2\sqrt{2}-1$, $\frac{18\sqrt{2}-15}{13}$) & $\frac{\ln(2)}{\ln(\frac{27}{13})}$ & $\frac{\ln(\frac{27}{13})}{\ln(2\sqrt{2}-1)}$ \\
      $I_1$ & (0, 0, $\sqrt{2}-1$) & ($2\sqrt{2}-1$, 0, 0) & $\frac{\ln(2)}{\ln(2\sqrt{2}-1)}$ &  \\
      $I_2$ & ($\sqrt{2}-1$, 1, $\sqrt{2}-1$) & ($2\sqrt{2}-1$, 0, 0) & $\frac{\ln(2)}{\ln(2\sqrt{2}-1)}$ &  \\
      $I_3$ & ($\sqrt{2}-1$, $3-2\sqrt{2}$, $5\sqrt{2}-7$) & ($2\sqrt{2}-1$, $29-20\sqrt{2}$, $71-50\sqrt{2}$) & $\frac{\ln(2)}{\ln(2\sqrt{2}-1)}$ & \\
      $U_1$ & ($0.27362, 0.27362, 0.18567$) & ($2.33597, 1.27297, 0.86747$) & $0.81698$ & $3.51527$ \\
      $U_2$ & (0.28801, 0.22580, 0.28801) & (2.33597, 1.27297, 0.86747) & 0.81698 & 3.51527 \\
\hline
\end{tabular}}
\end{center}
\end{table}

The nature of the remaining non-trivial fixed point $U_1$ and $U_2$ are quite distinct from the previously discussed. The stability analysis around these two fixed point reveals that they have the same set of three distinct eigenvalues, two of which are relevant. Therefore, they have the same critical exponents $\nu_T$ and $\phi$. These exponents do not correspond to any $q$-integer Potts universality class, as it has been verified in the case of the previously studied non-trivial fixed points. This peculiar behavior has been observed in a recent analysis of the 3AT model on a diamond-like hierarchical lattice \cite{pi2008}. As far as their location is concerned, $U_1$ lies on the merging line of the $\mathbf{P}$, $\mathbf{F}$ and $\mathbf{F}_1$ phases, while $U_2$ lies on a line where the merging of the $\mathbf{P}$, $\mathbf{F}$ and $\mathbf{F}_2$ phases occurs.

For the fixed points $I_1$, $I_2$, $I_3$, $P_4^{(1)}$, $P_4^{(2)}$, $P_4^{(3)}$ and $P_8$ we have identified  self-dual lines on which these points are located. This identification was crucial to determine the exact location for their respective coordinates as well as the associated critical exponents in closed form as presented in Table 2. However, in the case of both $U_1$ and $U_2$ we were not able to find out such identification of self-dual lines, and due to this fact we were limited to numerically determine the coordinates and critical exponents to these two points. In spite of this, a few words can be said about the points $U_1$ and $U_2$. Note that the self-dual planes corresponding to $\bar{t}_2 = \bar{t}_4$ and $\bar{t}_2 = \bar{t}_6$ shown in Figs.\ 4 and 5, respectively, are very similar to the findings reported in Ref.\ \cite{mtf1985}. This resemblance may indicate that $U_1$ and $U_2$ belong to one of the infinity universality classes that exists along the Baxter's line. The motivation for this conjecture comes from the fact that for the 2AT model we can write the specific heat exponent $\alpha$ along the Baxter's line as \cite{anjos}

\begin{equation}
\alpha={{2 - 2y}\over{3-2y}},
\end{equation}
where $y$ is given by
\begin{equation}
y={{2}\over{\pi}} {\cos}^{-1} \{ { {1} \over {2} } [\exp{( 8K_2)-1 ]}\}.
\end{equation}
We must remark that $K_2$ in Eq.\ (19) corresponds to the dimensionless four-spin coupling constant for the 2AT model, as defined in Ref.\ \cite{anjos}. In the Baxter's line $K_2$ is in the range [$0, {{1}\over{8}} \ln{(3)}$], and as a consequence $y$ is in the range $[0,1]$. One can represent the parameter $y$  along the Baxter's line as a function of the spatial dimension $d$ and correlation length critical exponent $\nu_T$ (once $d\nu_T=2-\alpha$), as follows \cite{anjos}
\begin{equation}
 y=\frac{3d\nu_T-4}{2d\nu_T-2}.
\end{equation}

By substituting the value $\nu_T = 0.816 \dots$ , obtained for $U_1$ and $U_2$ in the above expression we find for $d=2$ (corresponding to the square lattice) that $y \approx 0.711 \dots$ , whereas for $d_f = \ln 5/\ln 2$ (the Wheatstone bridge fractal dimension) we  get $y \approx 0.942 \dots$ .
Thus, in both $d=2$ and $d_f=\ln 5/\ln 2$ cases the exponent $\nu_T$ for $U_1$ and $U_2$ may belong to the Baxter's line. However, since we need two distinct critical exponents in order to determine the universality class we are left with a conjecture that should be probed by alternative methods.

As a final remark, it is worth mentioning that all self-dual curves and surfaces we have found along the present work are also presented in a recent treatment of the 3AT model on the diamond hierarchical lattice \cite{pi2008}.

\section{Conclusions}

In conclusion, we have studied the phase diagram and critical behavior of the 3AT model. We
have applied a Migdal-Kadanoff renormalization group technique in a hierarchical lattice recursively constructed from the Wheatstone bridge. Due
to the lattice structure we restrict ourselves to the ferromagnetic version of the model. Our results can be seen either
as an approximation for the 3AT on the square lattice, or as an exact result for the hierarchical lattice considered in the present work. We
have found the presence of four magnetic phases, which are listed in Table 1 along with their attractors. We believe that the phase
boundaries (the critical surfaces) obtained here represent a very good approximation to the exact results on the square lattice.
We have also found nine non-trivial fixed points, which are listed in Table 2 along with their critical exponents.
These nine non-trivial fixed points can be classified in four universality classes,
three of them are associated to the Potts model with $q = 2$ states ($I_1$, $I_2$ and $I_3$), $q = 4$ states ($P_4^{(1)}$,
$P_4^{(2)}$ and $P_4^{(3)}$) and $q = 8$ states ($P_8$). Moreover, there are two non-trivial fixed points ($U_1$ and $U_2$) which do not belong to the universality
class of the Potts model. We should remark that all those results are qualitatively similar to what was found in a recent work (see Ref.\ \cite{pi2008}). More interesting, however, is the analysis of the universality class associated to $U_1$ and $U_2$. Our numerical results show that the spatial dimension ($d=2$ or $d_f=\ln5/\ln2$) and correlation length critical exponent ($\nu_T =0.81698$) are compatible to one of the infinity universality classes that there are along the Baxter's line (see Eqs.\ 18-20). Therefore, we conjecture that both points $U_1$ and $U_2$ belong to
one of the infinity universality classes that exist along the Baxter's line.

\section*{Acknowledgements}
We would like to thank the Brazilian Research Council CNPq for partial financial support.

\newpage

\begin{figure}
\centering
\includegraphics[width=13cm]{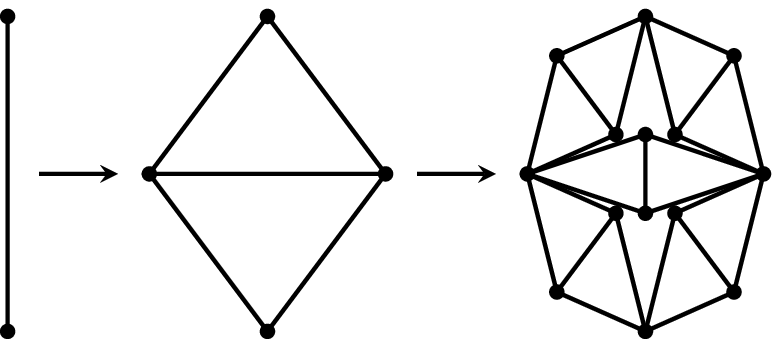}
\caption{The graph used for the generation of the hierarchical lattice considered in this work. The scaling factor $b=2$ and the number of bonds $m=5$ furnish a fractal dimension $d_f=\ln5/\ln2$.} \label{lattice}
\end{figure}

\begin{figure}
\centering
\includegraphics[width=13cm]{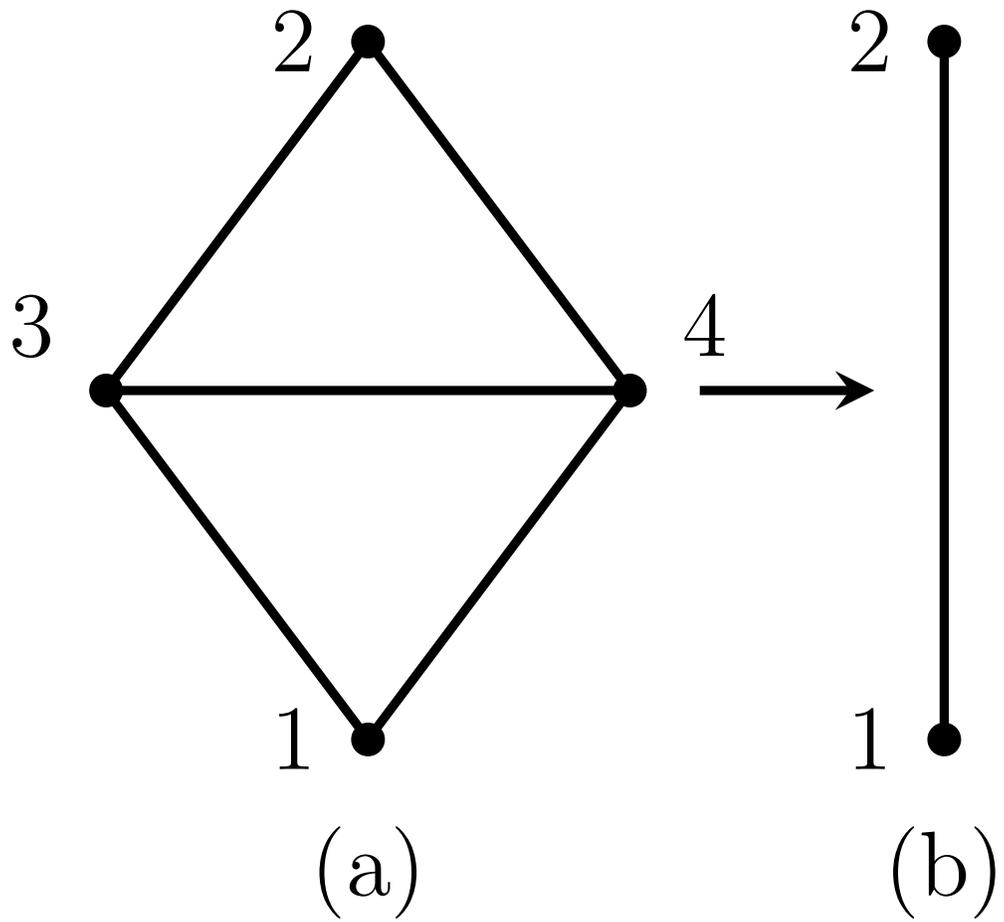}
\caption{Renormalization of the Wheatstone bridge hierarchical lattice.}
\end{figure}

\begin{figure} \centering
\includegraphics[width=13cm]{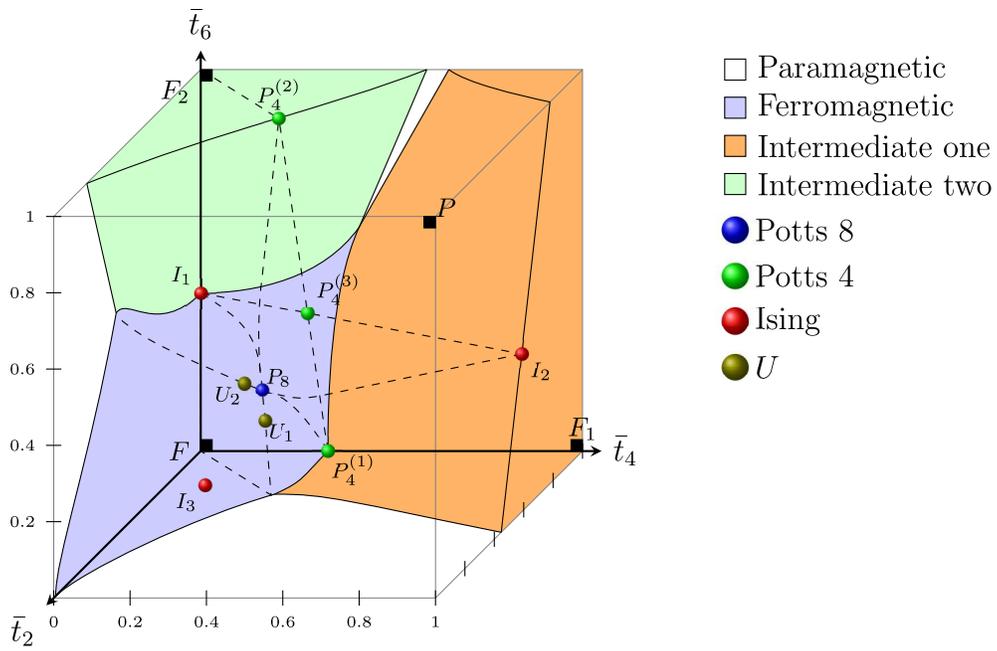}
\caption{Phase diagram of the 3AT model in ($\bar{t}_2, \bar{t}_4, \bar{t}_6$) space.}
\end{figure}

\begin{figure} \centering
\includegraphics[width=13cm]{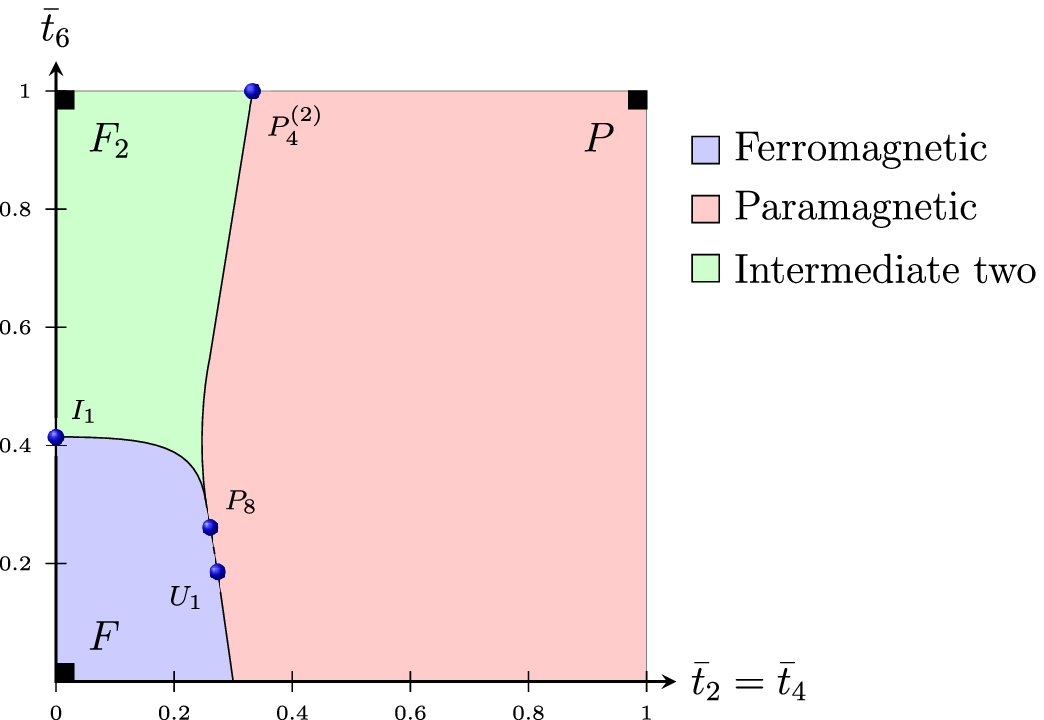}
\caption{Plane $\bar{t}_2=\bar{t}_4$ of Fig.\ 3.}
\end{figure}

\begin{figure} \centering
\includegraphics[width=13cm]{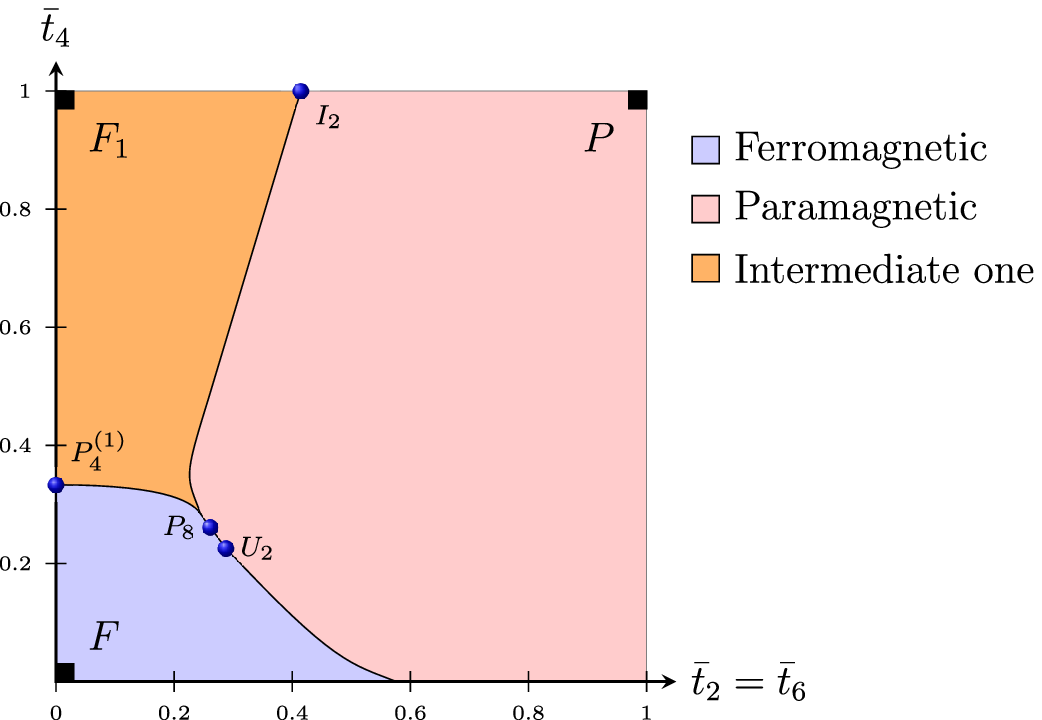}
\caption{Plane $\bar{t}_2=\bar{t}_6$ of Fig.\ 3.}
\end{figure}


\begin{thebibliography}{100}
\bibitem{gw1981} Grest G S and Widom M, 1981 \emph{Phys. Rev.} B \textbf{24} 6508
\bibitem{baxter} Baxter R J, 1982 \emph{Exactly Solved Models in Statistical Mechanics} (London: Academic Press)
\bibitem{gs1986} Goldschmidt Y Y, 1986 \emph{Phys. Rev. Lett.} \textbf{56} 1627
\bibitem{mf1988} Martins M J and de Felicio J D R, 1988 \emph{J. Phys. A: Math. Gen.} \textbf{21} 1117
\bibitem{rf1996} Fisch R, 1996 \emph{J. Appl. Phys.} \textbf{79}, 5088
\bibitem{fcc2003} de Felicio J R D, Chahine J, Caticha N, 2003 \emph{Physica} A \textbf{321} 529
\bibitem{pi2008} Piolho F A P, da Costa F A, Bezerra C G and Mariz A M, 2008 \emph{Physica} A \textbf{387} 1538
\bibitem{tm1996} Tsallis C and de Magalh\~aes A C N, 1996 \emph{Phys. Rep.} \textbf{268} 306
\bibitem{kg1981} Kaufman M and Griffiths R B, 1981 \emph{Phys. Rev.} B \textbf{24} 496
\bibitem{at1982} Alcaraz F C and Tsallis C, 1982 \emph{J. Phys A: Math. Gen.} \textbf{15}, 587
\bibitem{Ashkinteller} Ashkin J and Teller E, 1943 \textit{Phys. Rev.} \textbf{64} 178
\bibitem{fc1972} Fan C, 1972 \textit{Phys. Lett.} A \textbf{39} 136
\bibitem{mtf1985} Mariz A M, Tsallis C and Fulco P, 1985 \textit{Phys. Rev.} B \textbf{32} 6055
\bibitem{bmac2001} Bezerra C G, Mariz A M,  de Ara\'ujo J M and da Costa F A, 2001 \textit{Physica} A \textbf{292} 429
\bibitem{anjos} Anjos A S, Moreira D A, Mariz A M, Nobre F D and da Costa F A, 2007 \emph{Phys. Rev.} E \textbf{76} 041137
\end{thebibliography}
\end{document}